\documentclass[fleqn,11pt]{wlscirep}
\usepackage{multirow}
\usepackage{wrapfig}
\usepackage{float}
\usepackage{tabularx}%
\usepackage[title]{appendix}%
\usepackage{booktabs}%
\usepackage[T1]{fontenc}
\usepackage{makecell}%
\usepackage{graphicx}
\usepackage{subfig}
\usepackage{siunitx}
\usepackage{hyperref}
\usepackage{lineno}
\usepackage{changes}

\title{Unequal urban capacities for mobility adaptation under fuel-price shocks}
\author[1,+]{Zihao Zhang}
\author[2,+]{Yuanbo Zhang}
\author[2,3,*]{Xiaolei Ma}
\author[4,*]{Yuan Liao}
\affil[1]{Polytechnic University of Milan, Milan, Italy}
\affil[2]{School of Transportation Science and Engineering, Beihang University, Beijing, China}
\affil[3]{Key Laboratory of Intelligent Transportation Technology and System, Ministry of Education, Beijing, China}
\affil[4]{Department of Human Geography, Lund University, Lund, Sweden}
\affil[$*$]{Corresponding authors: \href{mailto:yuan.liao@keg.lu.se}{yuan.liao@keg.lu.se}; \href{mailto:xiaolei@buaa.edu.cn}{xiaolei@buaa.edu.cn}.}
\affil[$+$]{These authors contributed equally.}
\begin{abstract}
What a city makes reachable depends less on what it contains than on who can still afford to move when travel costs rise.
We leverage the 2026 US--Iran oil shock as a natural experiment, applying a hierarchical panel regression discontinuity design to 1.7 trillion point-of-interest visits across 122,000 neighbourhoods in China and the United States. 
Mobility range declined in nearly three-quarters of neighbourhoods, but responses varied systematically with pre-shock urban conditions. 
Exposure to energy-intensive travel explained the largest share of modelled heterogeneity in both countries, while adaptive capacity and activity composition further shaped how travel was reorganized. 
Longer baseline travel intensified contraction, whereas greater car dependence constrained adjustment. 
Crucially, similar mobility outcomes arose from different processes: some neighbourhoods maintained travel by absorbing higher costs, whereas others appeared structurally locked into travel they could not reorganize. 
Fuel-price shocks, therefore, act as urban stress tests, revealing which neighbourhoods a city keeps connected.
\end{abstract}
\usepackage{url}
\begin{document}
\keywords{energy vulnerability, urban form, fuel price shock, spatial heterogeneity, travel behavior}

\flushbottom
\maketitle

\thispagestyle{empty}

\section*{Introduction}
A city's transport infrastructure describes what is available; what residents can reach when the cost of moving rises describes what is accessible. 
Rising travel costs, from fuel markets, carbon pricing, or road charges, are the condition under which that difference becomes visible.
Whether a city can adapt depends on the exposure, the balance of essential and discretionary travel, substitution options, and financial slack that decide whether residents can reshape their travel or are locked into it; these together constitute a city's adaptive capacity~\cite{smit2006adaptation,gillingham2019tale,torne2024banning}.
This capacity is structurally embedded and unevenly distributed: it varies from one neighbourhood to another, shaped by urban form~\cite{ewing2010travel}, activity patterns~\cite{abbiasov202415}, and individual resources~\cite{alonso2023transport}, so a mere city-level average conceals who can adapt and who cannot. \par

Observing the heterogeneity of who adapts and who cannot requires observing behaviour at a fine spatial scale across a wide range of urban systems, a resolution that travel surveys and aggregate statistics rarely achieve.
Large-scale mobility data now make it possible to trace movement between neighbourhoods and destinations for entire national populations~\cite{chang2021mobility,nilforoshan2023human,pappalardo2023future} and measuring how those populations adjust their travel when the cost of moving changes~\cite{liao2025uncovering}.
They turn the adaptive capacity of mobility systems from a concept into a measurable property of neighbourhoods and reveal that a rise in the cost of moving is borne far more unevenly than the average suggests. \par

Current research lacks causal evidence on how neighbourhoods in varying urban systems differ in their ability to adjust to higher travel costs.
Transport poverty research identifies which households are exposed on these terms, documenting the disproportionate burdens on low-income and car-dependent households as fuel prices rise~\cite{mattioli2016transport,mattioli2018vulnerability}.
Such literature establishes who is vulnerable, not how they adjust.
Fuel-demand studies measure adjustment directly, yet estimate it in aggregate as reductions in driving that follow price increases~\cite{goodwin2004elasticities,graham2002demand,gillingham2014identifying}, obscuring the local variation through which adaptive capacity becomes visible.
Moreover, the existing literature largely relies on evidence from individual countries, which cannot show whether the findings persist across urban systems that transmit global fuel-price shocks differently.
The meaning of behavioural response under varying adaptive capacity, therefore, remains ambiguous: a reduced mobility range may represent flexible substitution or forced curtailment, whereas little change may reflect either financial buffering or structural lock-in.
Resolving this ambiguity requires following a shared external shock across different urban systems~\cite{hong2021measuring}, at sufficient spatial resolution to separate global averages from the local conditions that enable or constrain adaptation~\cite{smit2006adaptation}. \par

The escalating US--Iran conflict disrupted global oil supplies, pushing the international crude benchmark up by roughly 90\% within ten weeks and sharply increasing the cost of daily mobility worldwide across diverse neighbourhoods, creating a natural experiment: an external fuel-price shock.
We leverage this shared shock in the urban systems in China and the US, where policy-guided and market-based pricing transmitted it to travellers with different intensity.
Using mobility data covering approximately one trillion POI visits across 80,000 US census block groups (CBGs) and 650 billion visits across 40,000 Chinese subdistricts (jiedao), we apply a hierarchical panel regression discontinuity design~\cite{raudenbush2015multisite, rhoads2016optimal,wellenius2021impacts} to estimate the shock-induced change in mobility range for each neighbourhood.
We define \textit{mobility rigidity} as the extent to which a neighbourhood maintains its mobility range as travel becomes more costly, without assuming that such persistence necessarily indicates resilience~\cite{mattioli2018vulnerability}.
Instead, we interpret each response through exposure, institutional price transmission, and three structural dimensions: sensitivity (daily activity structure), the absorptive capacity to bear higher costs, and the substitutive capacity to reorganize trips~\cite{smit2006adaptation}.
Linking shock responses to these pre-shock conditions distinguishes flexible reorganization from forced curtailment, and financial absorption from structural lock-in.
The shock thereby converts each neighbourhood's response into a causal readout of its pre-shock adaptive capacity. \par

Here, we show that widespread travel contraction masks a geography of adaptive capacity.
Mobility range fell in nearly three-quarters of neighbourhoods in both countries, but exposure (car ownership and pre-shock travel distance) and income explained this variation more strongly than transit, density, or land-use mix.
Crucially, maintained travel arose both where higher costs could be absorbed and where daily travel could not be reorganized.
Fuel-price shocks act as urban stress tests, revealing an adaptive architecture through which a uniform rise in travel costs becomes an uneven loss of access. \par

\section*{Results}
\subsection*{Defining mobility rigidity with two urban systems under one fuel-price shock}
\begin{figure}[!ht]
\centering
\includegraphics[width=1\textwidth]{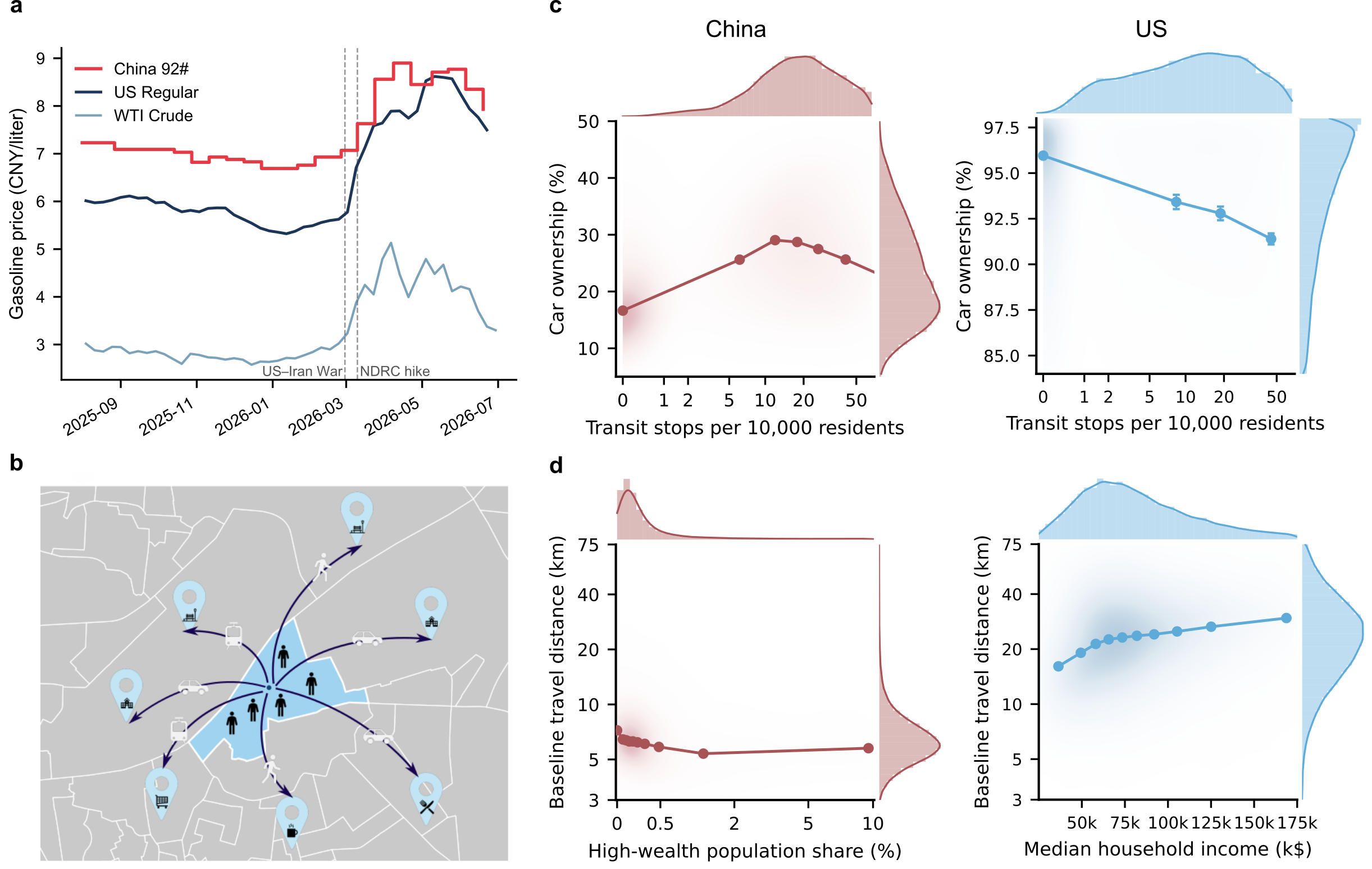}
\caption{\textbf{Measuring mobility responses to an exogenous fuel-price shock in China and the US.}
\textbf{a}, Retail gasoline and crude benchmark prices (CNY/litre), Sep 2025--Jul 2026. 
Dashed lines mark the onset of the US--Iran war and the subsequent NDRC-administered price hike, which jointly delimit the shock window applied in the panel RDD.
\textbf{b}, Schematic of the activity-space unit measured: a focal neighbourhood in blue and its observed travel flows to various destinations (POIs), reached by walking, transit, and car (transport mode information only available in China's data).
\textbf{c}, Joint distribution of car ownership (\%) against transit supply (transit stops per 10,000 residents, log axis) for China (left, red) and the US (right, blue). 
Background: 2D Gaussian-KDE density of all spatial units; foreground: binned medians with bootstrap-SE error bars; marginal KDEs on the top and right axes.
\textbf{d}, Joint distribution of baseline travel distance (km, log axis) against the socioeconomic gradient, proxied by high-wealth population share (\%, log axis) for China (left, red) and median household income (k\$) for the US (right, blue). 
Density, binning, and marginal conventions as in \textbf{c}.}
\label{fig:fig1}
\end{figure}

The escalation of the US--Iran conflict pushed the international crude benchmark up by 90\% within 10 weeks (Fig.~\ref{fig:fig1}a).
The same signal reached travellers unevenly.
US retail gasoline prices track crude markets and absorb spikes within days, whereas Chinese retail prices are guided by the National Development and Reform Commission (NDRC) regularly, thereby delaying and dampening price transmission.
We summarise this institutional filtering as a price-transmission ratio, PT --- the percentage change in domestic retail fuel prices divided by the percentage change in international crude (see the Methods section).
PT reached 58\% in the US but only 36\% in China (Supplementary Fig.~\ref{fig:si_oil_price}). \par

One global shock thus arrived as two magnitudes of local cost, raising the question of how neighbourhoods in each urban system adapted.
A neighbourhood that barely changes its travel may be affluent enough to absorb the higher cost without adjusting, or bound to trips it cannot drop.
Observed change therefore carries no information about vulnerability until the conditions that produced it are accounted for.
We define \textit{mobility rigidity} as the extent to which a neighbourhood maintains its mobility range as travel costs rise, a property that is neither protective nor harmful by construction (see ``Mobility rigidity'' in the Methods section).
We measure it from 651 billion POI visits across 39,538 subdistricts in China and 1,046 billion visits across 82,908 census block groups in the US (Fig.~\ref{fig:fig1}b), quantifying how far residents travel and which destinations they reach before and after the shock. \par

Vulnerability frameworks assume that exposure, sensitivity, and adaptive capacity interact, prompting us to frame rigidity as two questions.
How much cost lands on a neighbourhood is set by its \textit{exposure} --- pre-shock reliance on energy-intensive travel, measured as average baseline mobility distance and vehicle ownership --- and by the national price transmission that scales the global shock of crude oil.
Whether that cost can be shed depends on three further properties: \textit{sensitivity}, the share of essential vs. discretionary trips, which characterise how much travel is available to cut; \textit{absorptive capacity}, the financial room to keep moving and simply pay more, indexed by income, and \textit{substitutive capacity}, the option to shift onto transit or toward closer destinations, indexed by transit density and land-use diversity.
\begin{equation}
\text{Rigidity} = f\!\left(\text{Exposure},\; \text{PT},\; \text{Sensitivity},\; \text{Absorptive capacity},\; \text{Substitutive capacity}\right).
\label{eq:rigidity}
\end{equation}

China and the US place these dimensions in different relations to one another (Fig.~\ref{fig:fig1}c,d).
US baseline distance climbs with median household income, coupling exposure to absorptive capacity, so the places most committed to long trips are also the best cushioned against their cost.
In China, baseline distance stays flat across the wealth gradient, and the two are decoupled.
Motorization also behaves differently in the two systems: the US car--transit trade-off makes vehicle access an inverse proxy for substitution, whereas China's hump-shaped pattern, in which ownership and transit supply co-evolve through urbanization, breaks that mapping.
The same rigidity framework, therefore, meets two contrasting urban systems.
We first estimate the shock's average effect on mobility range, then reveal which neighbourhoods adapted and which could not.

\subsection*{A pervasive contraction of mobility range unevenly borne across neighbourhoods}

The fuel-price shock produced a clear contraction of mobility range in both countries.
Operationalizing the concept of mobility rigidity, we recover the discontinuity estimate $\hat{\tau}_i$ for every neighbourhood $i$ from a hierarchical panel regression discontinuity design centred on each country's price breakpoint, 28 February 2026 for the US and 10 March 2026 for China (see the section ``Quantifying mobility rigidity'' of the Methods).
Averaged across neighbourhoods, mobility distance fell by 2.33\% in China (95\% CI [$-2.42$\%, $-2.24$\%]) and 19.0\% in the US (95\% CI [$-24.9$\%, $-12.6$\%]), a gap of 5.1 times even after scaling by each country's crude-to-retail price transmission (Fig.~\ref{fig:fig1}a and Supplementary Table~1). 
These correspond to implied mobility--price elasticities of approximately $-0.50$ in the US and $-0.08$ in China, consistent with published estimates for the US and below the published range for China, for reasons detailed in Supplementary Note ``External validation of estimated shock effects''. \par

Across alternative bandwidths, China's discontinuity estimate is reproduced almost exactly across bandwidths of six to eight weeks (-2.33\% to -2.35\%), while the US contraction remains large from two to four weeks (-23.6\% to -19.0\%) and reaches -10.3\% at eight weeks.
The isolated six-week estimate is near zero because this window's pre-period extends back to the late-January fuel price run-up, when travel was already adjusting; part of the contraction is therefore absorbed into the pre-shock baseline rather than attributed to the post-shock response. 
At eight weeks, the wider window re-anchors the baseline in the pre-run-up period and extends further into the post-shock period, restoring a negative estimate.
Placebo breakpoints were set one year earlier in 2025, yielding effects of only 0.1--0.6\% in the US and -0.3\% to +0.1\% in China, much smaller than the shock estimates.
These small placebo estimates across bandwidths support attributing the observed contraction to the fuel-price shock rather than recurring spring seasonality or other time-varying factors (see Supplementary Tables~3 and 4). \par

\begin{figure}[!ht]
\centering
\includegraphics[width=0.9\textwidth]{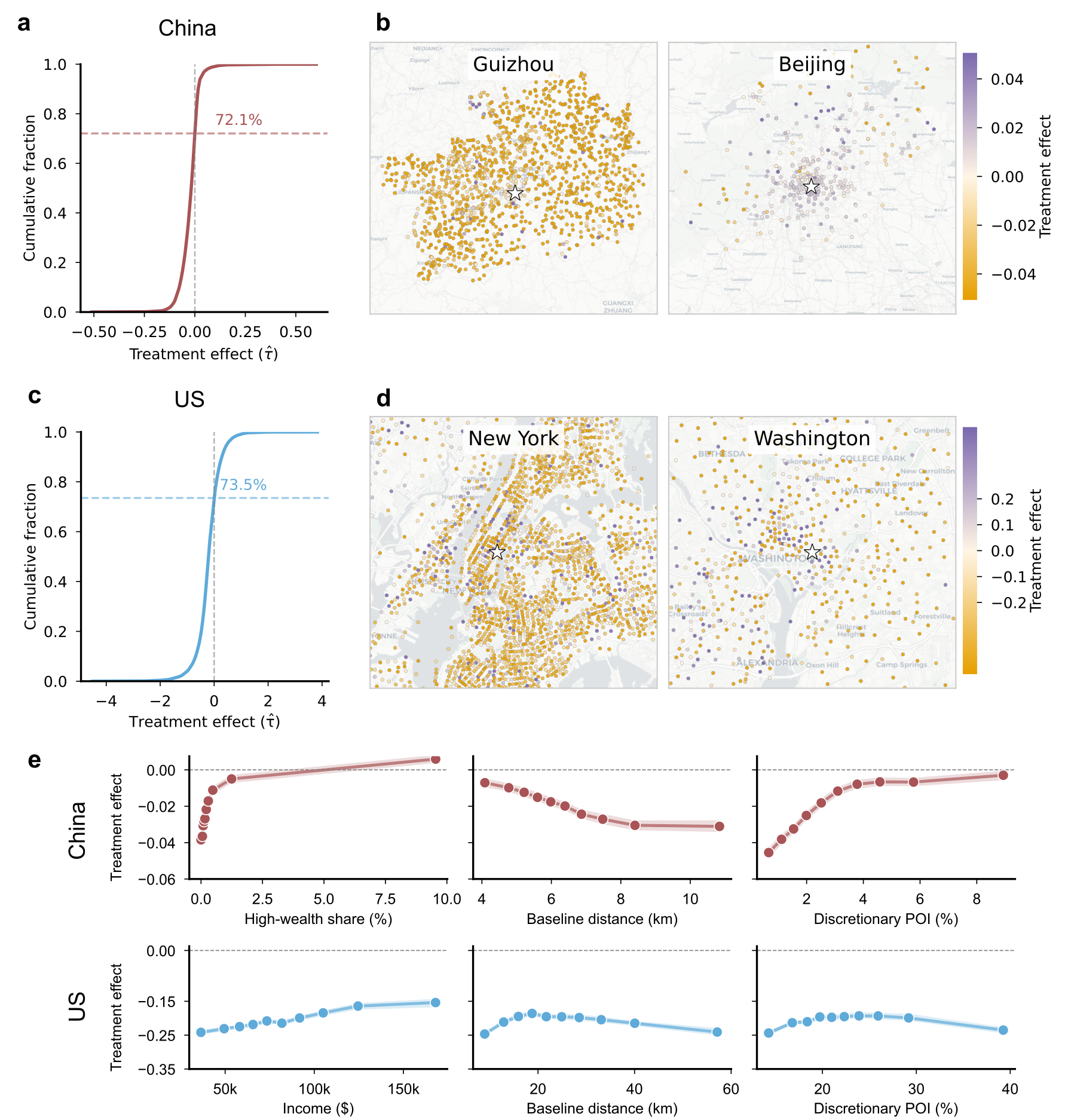}
\caption{\textbf{A widespread reduction in mobility range, unevenly distributed across neighbourhoods.}
\textbf{a}, Cumulative distribution of neighbourhood-level treatment effects ($\hat{\tau_i}$) for China.
The dashed vertical line marks $\hat{\tau_i}=0$.
\textbf{b}, Spatial distribution of $\hat{\tau_i}$ for two distinct Chinese regions, Guizhou (left) and Beijing (right); points are neighbourhoods coloured from reductions (orange) to increases (purple), with the regional centre marked by a star.
\textbf{c}, Cumulative distribution of neighbourhood-level treatment effects for the US, as in \textbf{a}.
\textbf{d}, Spatial distribution of $\hat{\tau_i}$ for two representative US cities, New York (left) and Washington (right); points represent neighbourhoods, coloured as in \textbf{b}.
\textbf{e}, Binned treatment effect against three pre-shock rigidity dimensions for China (top, red) and the US (bottom, blue): the socioeconomic gradient (high-wealth population share, \%, for China; median household income, k\$, for the US), baseline mobility distance (km), and discretionary POI share (\%). 
Points are binned medians; shaded bands are bootstrap standard errors of the binned median.}
\label{fig:fig2}
\end{figure}

The contraction was widespread, with mobility range falling in 72.1\% of Chinese and 73.5\% of US neighbourhoods (Fig.~\ref{fig:fig2}a,c), and the effects were more dispersed in the US.
Responses varied sharply in different places (Fig.~\ref{fig:fig2}b,d).
However, unequal capacity to adapt is primarily a condition within cities rather than between them.
Decomposing $\hat{\tau}_i$ into between- and within-city components attributes 16.6\% of the variance in the US to differences between cities and 83.4\% to differences among neighbourhoods of the same city; in China the corresponding shares are 33.0\% and 67.0\%.
Neighbourhoods of the same city therefore diverge more from one another than cities do among themselves, in both urban systems.
The between-city share is twice as large in China, where cities span a wider development gradient than in the US (Fig.~\ref{fig:fig1}c). \par

Mobility contractions follow the pre-shock rigidity dimensions (Fig.~\ref{fig:fig2}e).
Longer baseline distances are associated with larger reductions in China, whereas the relationship in the US is non-monotonic.
In both countries, reductions weakened with affluence and with the share of discretionary destinations in people's daily activities (see the covered POI categories in Supplementary Table~5).
Because the hierarchical model shrinks noisier neighbourhood-level estimates toward the national mean, the observed socioeconomic gradients in Fig.~\ref{fig:fig2}e could in principle arise if shrinkage were systematically stronger in poorer neighbourhoods. 
Re-estimating neighbourhood effects without shrinkage reproduces the same gradients, ruling out this explanation (Supplementary Tables~6 and 7).
China's transport mode change patterns suggest that the quantified contraction in mobility range reflects a genuine travel response rather than a measurement artifact (Supplementary Tables~8 and 9).
In China, trip frequency and driving share both declined, while the non-motorized share rose, and the neighbourhoods with the strongest declines in driving were those with the largest gains in non-motorized travel, a pattern of short-trip substitution and outright trip suppression consistent with behavioural response to fuel-price increase (Supplementary Tables~8 and 9).
Because exposure, sensitivity, absorptive capacity, and substitutive capacity covary, we subsequently model these dimensions jointly to explain the observed heterogeneous contraction in mobility range. \par

\subsection*{Disentangling the structural dimensions of mobility rigidity}
\begin{figure}[!ht]
\centering
\includegraphics[width=0.9\textwidth]{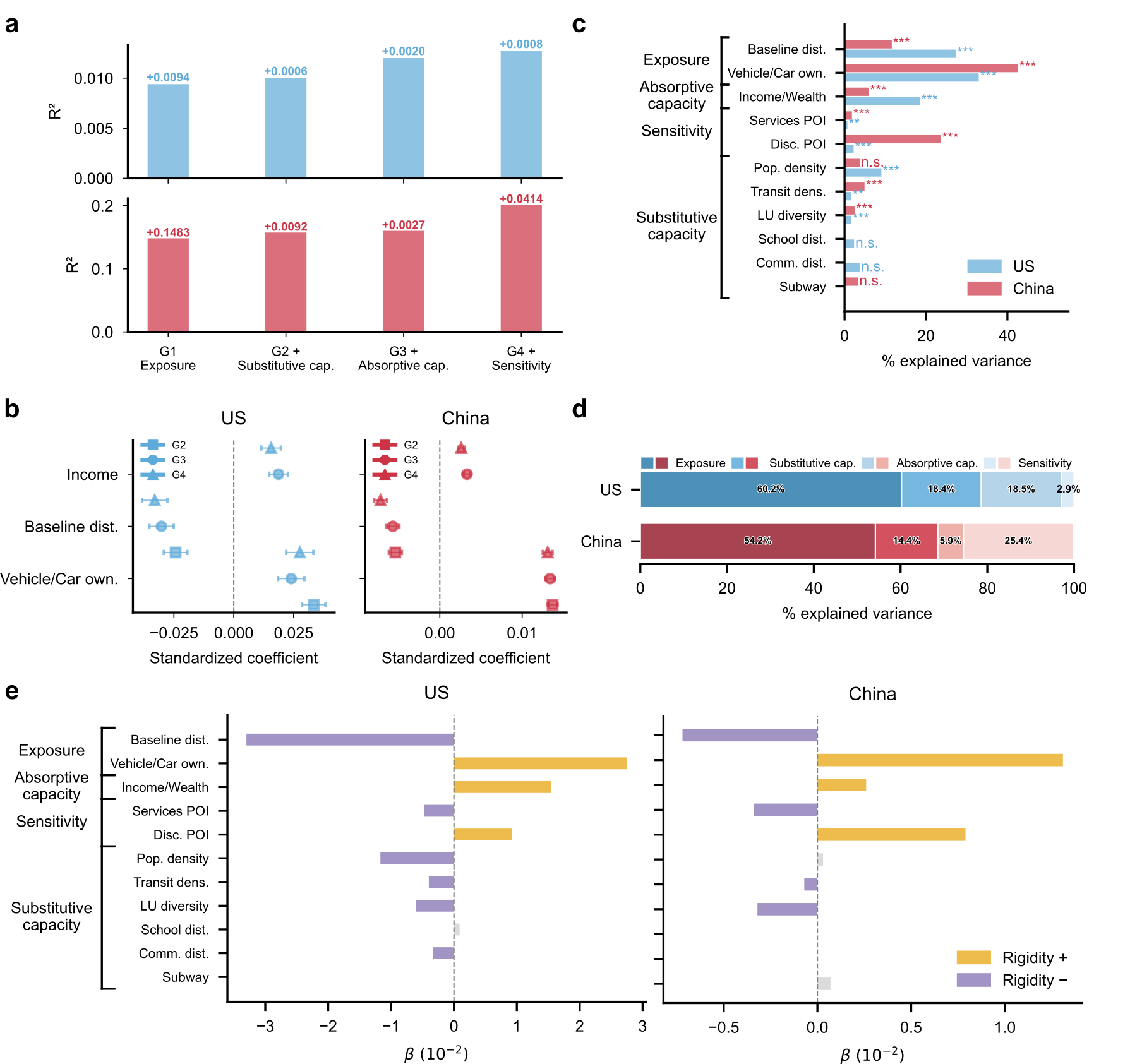}
\caption{\textbf{Decomposing mobility rigidity into its structural dimensions.}
\textbf{a}, Cumulative explanatory power ($R^2$) of four OLS models of the neighbourhood-level treatment effect $\hat{\tau}$ for the US (top, blue) and China (bottom, red); bar heights are cumulative $R^2$ and annotations give the increment contributed by each group.
\textbf{b}, Standardized coefficients of three predictors (income/wealth, baseline distance, vehicle ownership) persist across the progressive models; error bars are 95\% confidence intervals, and the dashed line marks $\beta=0$.
\textbf{c}, Variable-level LMG decomposition of the full-model (G4) $R^2$, expressed as the percentage of explained variance attributable to each predictor and grouped by rigidity dimension ($^{***}p<0.001$, $^{**}p<0.01$, n.s. $p\geq0.05$).
\textbf{d}, Dimension-level aggregation of the LMG values in \textbf{c}; shares sum to 100\% within each country. 
\textbf{e}, Standardized full model coefficients ($\beta$, $\times 10^{-2}$) for all predictors, grouped by rigidity dimension. 
Because $\hat{\tau}$ is predominantly negative, positive coefficients denote a weaker mobility reduction and are coloured as rigidity-increasing (yellow); negative coefficients denote a stronger reduction and are coloured as rigidity-decreasing (purple); grey bars are not significant at $p<0.05$.}
\label{fig:fig3}
\end{figure}

The dimensions of mobility rigidity are conceptually distinct but empirically entangled.
We answer whether they carry separate information in two steps.
A progressive series of four regressions of the treatment effect $\hat{\tau_i}$ on pre-shock characteristics adds one dimension at a time in framework order --- exposure (G1), substitutive capacity (G2), absorptive capacity (G3), sensitivity (G4) --- so that each increment in $R^2$ measures what a dimension adds once all preceding ones are held fixed (Fig.~\ref{fig:fig3}a; the section ``Modelling the structural dimensions of mobility rigidity'' of the Methods, and Supplementary Table~10 for collinearity diagnostics).
Because that increment depends on entry order, we then decompose the full-model $R^2$ using the Lindeman--Merenda--Gold method, which averages each predictor's contribution over all possible entry orders and yields shares that sum exactly to the model's explanatory power (see the section ``Modelling the structural dimensions of mobility rigidity'' of the Methods). \par

Observable pre-shock structure organizes mobility rigidity in China but leaves it largely unexplained in the US (raw $R^2 = 0.20$ versus $0.013$; Fig.~\ref{fig:fig3}a).
This asymmetry is partly consistent with China's wide development gradient (Fig.~\ref{fig:fig1}c) and weak price transmission (PT $=0.36$ versus $0.58$; Fig.~\ref{fig:fig1}a), which delivers the shock to neighbourhoods whose structural conditions differ sharply, and those conditions predict which of them contract.
While in the US, high motorization and a larger increase in travel costs may compress structural variation, leaving responses tied to household circumstances that the modelled variables do not capture.
Estimation noise in $\hat{\tau}_i$ imposes a further ceiling on attainable fit in the US (reliability ratio $\lambda=0.66$) compared to a lower noise level in China ($\lambda=0.82$). \par

The explanatory gains of structural dimensions on mobility contraction arise primarily from exposure, followed by absorptive capacity and sensitivity in the US and sensitivity and substitutive capacity in China (Fig.~\ref{fig:fig3}a).
Exposure contributes the most in either country ($\Delta R^2 = 0.1483$ in China and $0.0094$ in the US), whereas the other dimensions account for about a quarter of the explanatory power on this entry-order basis, led in China by sensitivity ($+0.0414$), substitutive capacity ($+0.0092$), and absorptive capacity ($+0.0027$), and in the US by absorptive capacity ($+0.002$), sensitivity ($+0.0008$), and substitutive capacity ($+0.0006$).
Exposure determines the scale of travel subject to the shock, whereas the remaining dimensions determine whether that travel can be reorganized.
In summary, we highlight three findings that follow.
First, exposure is dominant in both countries; the secondary organizing dimensions differ: sensitivity and substitutive capacity in China, absorptive capacity and sensitivity in the US.
Second, neighbourhood activity composition systematically differentiates behavioural responses in both countries.
Third, similar behavioural outcomes can emerge from fundamentally different adaptive processes.

\subsection*{Exposure creates pressure to adapt, but automobile dependence constrains adjustment}

Exposure dimension provides the strongest initial organization of mobility rigidity in both countries.
Baseline distance and vehicle ownership alone explain \(R^2=0.0094\) in the US and \(R^2=0.1483\) in China, compared with full-model values of 0.013 and 0.20, respectively (Supplementary Tables 11 and 12).
The order-independent decomposition confirms this dominance: exposure accounts for 60.2\% of explained variation in the US and 54.2\% in China (Figs.~\ref{fig:fig3}c,d). \par

The two forms of exposure, however, operate in opposite directions.
In the exposure-only model, longer baseline distance predicts a stronger mobility contraction in both the US and China (\(\beta_{\mathrm{US}}=-0.0276\), \(\beta_{\mathrm{China}}=-0.0053\); both \(p<0.001\)), whereas greater vehicle ownership predicts a weaker contraction (\(\beta_{\mathrm{US}}=+0.0352\), \(\beta_{\mathrm{China}}=+0.0138\); both \(p<0.001\)).
These opposing associations remain after accounting for substitutive capacity, absorptive capacity, and sensitivity (G4: baseline distance, \(\beta_{\mathrm{US}}=-0.0330\), \(\beta_{\mathrm{China}}=-0.0072\); vehicle ownership, \(\beta_{\mathrm{US}}=+0.0276\), \(\beta_{\mathrm{China}}=+0.0131\); all \(p<0.001\); Figs.~\ref{fig:fig3}b,e). \par

Exposure, therefore, does not translate directly into behavioural change.
Longer travel creates greater cost pressure and more distance to reduce, whereas automobile dependence ties everyday activities to a mode that is difficult to replace.
Fuel-price shocks are amplified by the spatial extent of travel but attenuated by the car dependence through which that travel is organized.
Fuel-price shocks, therefore, expose an inequality that urban infrastructure alone cannot resolve: alternatives shape how travel can be reorganized, whereas car dependence determines whether the rising cost of travel can be borne without curtailment. \par

\subsection*{Neighbourhoods with more discretionary travel resist the mobility range contraction}
\begin{figure}[!ht]
\centering
\includegraphics[width=1\textwidth]{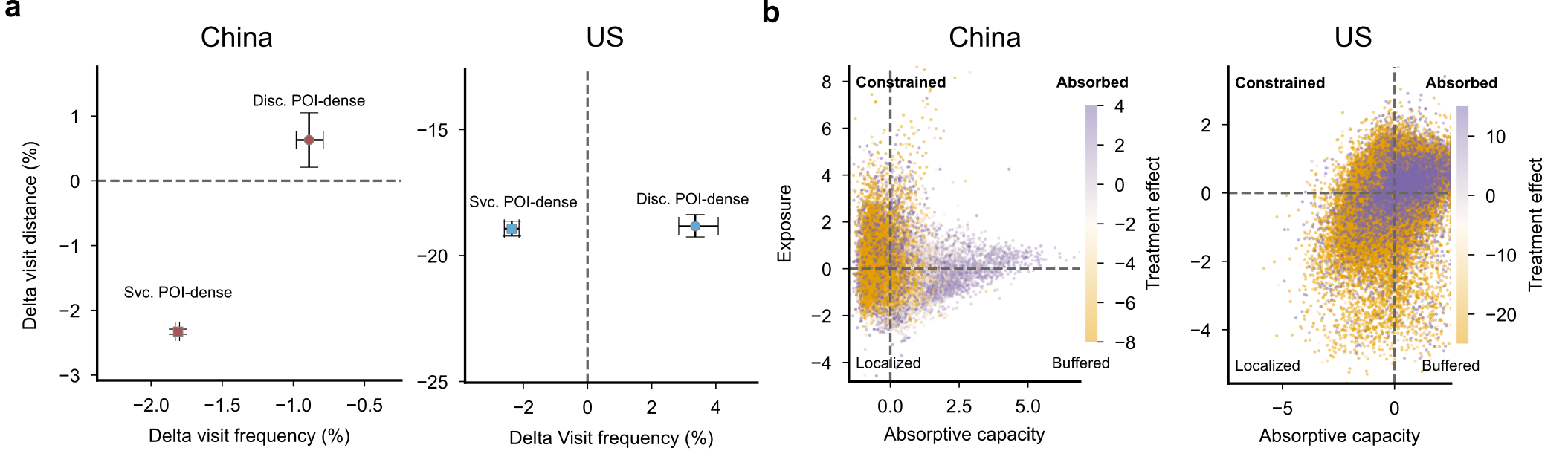}
\caption{\textbf{Explaining mobility rigidity: discretionary substitution and the divergent origins of rigidity.}
\textbf{a}, Post-shock change in visit distance (\%) against the post-shock change in visit frequency (\%) for discretionary-POI-dense and service-POI-dense neighbourhoods, in China (left) and the US (right); points are group medians and bars are bootstrap median errors.
\textbf{b}, Every neighbourhood placed by its pre-shock absorptive capacity ($x$) and exposure ($y$) as z-scores, coloured by treatment effect ($\hat{\tau}$; orange, stronger reduction, to purple, weaker or positive), for China (left) and the US (right). 
$x$ is the z-score of log median household income; $y$ is the equally weighted average of the z-scores of pre-shock baseline travel distance and vehicle ownership.}
\label{fig:fig4}
\end{figure}

A smaller range contraction, where daily activities lean more toward the discretionary, could mean trips were dropped rather than shortened.
To test whether the overall reduction in mobility range is partly driven by reduced frequency, we classify neighbourhoods by their pre-shock POI composition into discretionary-dominated (dining, accommodation, sports, and scenic spots) and services-dominated (education, healthcare, government, corporate services, finance, and personal services) neighbourhoods.
A discretionary-dominated neighbourhood is one where the share of discretionary POI visits exceeds that of service-related activities.
Then we compare the fuel-price shock effect and weekly visit frequency across these two types (Fig.~\ref{fig:fig4}a). \par

In the US, adaptation runs mainly through spatial substitution.
Discretionary-dominated neighbourhoods contract travel distance by $-18.8\%$ (95\% CI [$-19.3$, $-18.4$]) while visit frequency increases ($+3.4\%$, 95\% CI [$+2.8$, $+4.0$]); services-dominated neighbourhoods contract distance by $-18.9\%$ (95\% CI [$-19.2$, $-18.6$]) with a modest frequency decline ($-2.4\%$, 95\% CI [$-2.6$, $-2.1$]).
In China, adaptation runs mainly through trip forgoing.
Discretionary-dominated neighbourhoods maintain travel distance ($+0.63\%$, 95\% CI [$+0.21$, $+1.05$]) while visit frequency falls ($-0.89\%$, 95\% CI [$-0.98$, $-0.79$]); services-dominated neighbourhoods contract on both margins ($-2.33\%$ distance, $-1.81\%$ frequency). \par

Both countries share one regularity across these divergent mechanisms: discretionary-dominated neighbourhoods resist overall mobility range contraction more strongly than services-dominated ones, absorbing the shock through frequency in China and through closer substitutes in the US.
This reflects the association between a neighbourhood's pre-shock lifestyle (as reflected in activity composition) and its mobility rigidity.

\subsection*{The divergent origins of mobility rigidity}

Observed mobility change becomes interpretable only when read against the exposure and adaptive capacity that produced it. 
We therefore position each neighbourhood by its pre-shock absorptive capacity and exposure and overlay its local treatment effect (Fig.~\ref{fig:fig4}b). 
The resulting quadrants distinguish four structural profiles. 
\emph{Absorbed} neighbourhoods combine high exposure with the resources to maintain long-distance travel, whereas \emph{constrained} neighbourhoods face similarly high exposure with limited financial capacity. 
\emph{Localized} neighbourhoods have low exposure and limited resources, leaving little distance to shed; \emph{buffered} neighbourhoods combine low exposure with high absorptive capacity. \par

The origins of contraction differ sharply between the two urban systems. 
In the US, 33.4\% of contracting neighbourhoods fall in the \emph{absorbed} profile, compared with 24.1\% in the \emph{constrained} profile. 
Together with the predominance of spatial substitution above, this pattern points to capacity-enabled reorganization. 
In China, by contrast, 37.2\% of contracting neighbourhoods are \emph{constrained}, compared with only 11.0\% \emph{absorbed}. 
Within the constrained profile in China, 82.7\% of neighbourhoods contract, with mobility range falling by 3.63\%—more than four times the 0.86\% reduction among absorbed neighbourhoods. 
Contraction is thus concentrated where long-distance exposure meets limited resources, consistent with constraint-driven curtailment rather than flexible reorganization. \par

Together, the neighbourhood-scale discontinuity estimates and pre-shock capacity measures make the latent adaptive architecture of urban systems observable. 
They reveal not only where a common fuel-price shock changes travel, but whether higher costs are absorbed through financial resources, accommodated through spatial reorganization, or borne through constrained reductions in mobility. 
The opposing patterns in the US and China show that these pathways are structured by local inequalities and different mobility systems rather than by the price shock alone. 
Mobility rigidity, therefore, captures how a common external shock is translated into unequal urban capacities to adapt. \par

\section*{Discussion}
Our work interprets neighbourhood-level response heterogeneity as evidence of unequal adaptive capacity, extending research on adaptation and fuel-price vulnerability~\cite{smit2006adaptation,mattioli2016transport,mattioli2018vulnerability}.  
An external cost shock makes this latent property observable by revealing differences in substitution options, financial buffers, and activity constraints that remain hidden under stable travel costs.  
Mobility rigidity therefore emerges through a cascade from institutional price transmission to travel costs and behavioural adjustment, shaped jointly by exposure, sensitivity, and adaptive capacity. \par

Behavioural response, however, is not equivalent to welfare loss.  
Reduced mobility may reflect either effective substitution or sacrificed activities, while maintained mobility may indicate either financial buffering or structural dependence on travel that cannot be reorganized (Fig.~\ref{fig:fig4}).  
The relevant distinction is therefore not how much travel changes, but whether residents retain access to essential activities and feasible alternatives~\cite{alonso2023transport,pereira2017distributive}.  
Read alongside pre-shock conditions, mobility responses reveal whether higher costs are absorbed, accommodated through substitution, or borne through constrained curtailment. \par

Our results reposition mobility rigidity as a relational property of urban systems rather than a simple function of how far people travel. 
Exposure, particularly car dependence, shapes whether higher costs can be absorbed, while built form and accessibility matter by determining whether feasible substitutes exist; their effects are therefore conditional rather than independent~\cite{smit2006adaptation,mattioli2018vulnerability,geurs_accessibility_2004,ewing2010travel}. 
This ordering is not specific to cost shocks: across five global cities, activity spaces account for more variation in place exposure than demographics or transit proximity~\cite{fan2026hidden}.
Activity-space composition provides a further organizing mechanism: discretionary-oriented neighbourhoods preserved their mobility range more strongly, whereas service-oriented neighbourhoods contracted more, showing that adaptive responses depend on the organization of everyday activities rather than on a simple distinction between flexible and essential trips~\cite{yang2023identifying}.
Therefore, transport disadvantage arises from the interaction between required mobility, available alternatives, and the resources needed to use them—not from travel reduction alone~\cite{mattioli2016transport,alonso2023transport,pereira2017distributive,liao2026space}. \par

The US--China comparison shows that the same fuel-price shock was filtered through different mobility systems.
Measured dimensions of mobility rigidity accounted for far more variation in China (reliability-corrected $R^2$ = 0.25) than in the US (reliability-corrected $R^2$ = 0.019). 
A relatively low $R^2$ in our OLS models is expected because the dependent variable is an estimated treatment effect rather than a directly observed outcome. 
First-stage estimation noise reduces the apparent fit~\cite{pagan1984econometric}, while much of the remaining variation in treatment effects is inherently idiosyncratic even when systematic patterns are present~\cite{raudenbush2015learning,ding2019decomposing}.
This gap in model performance suggests that adaptation is more closely tied to observable neighbourhood conditions in China than in the US, where responses may depend more on unmeasured constraints such as employment schedules and household finances, calling for further study.  
Despite this difference in explanatory power, all the significant associations were consistent in direction across the two countries in the full specification (G4), strengthening the evidence that mobility rigidity is systematically related to exposure, sensitivity, and adaptive capacity at the neighbourhood level. \par

Our findings show that policies raising the cost of travel should be judged by whose access they preserve, not simply by how much travel falls.
Fuel taxes, road pricing, and related measures are commonly evaluated through average demand reductions and household expenditure burdens~\cite{creutzig2015transport,torne2024banning}. 
Such measures can obscure whether reduced travel reflects flexible reorganization or constrained curtailment, with very different consequences for access to everyday activities. 
Policies that raise travel costs should therefore be paired with accessible destinations, credible transport alternatives, and targeted support where essential travel cannot readily be reorganized. \par

Our analysis observes neighbourhood responses in the weeks following a single shock. 
It therefore cannot show how individual well-being changed, how responses differed within neighbourhoods, or how households and cities might adapt over longer periods through relocation, vehicle replacement, or land-use change. 
Research following prolonged or repeated shocks could test whether the same effects persist and, crucially, whether mobility adjustments preserve access to essential activities. 
Despite these limits, our work shows that the shock serves as an urban stress test, making cities' adaptive capacities visible. 
Mobility under stress thus reveals the unequal capacity to adapt that an equitable transition to low-carbon transport depends on. \par

\section*{Methods}
In the Methods section, we first describe the datasets used (see ``Datasets''), followed by the derived mobility rigidity and its structural determinants (see ``Mobility rigidity'').
We then describe the hierarchical regression discontinuity design (RDD) used to quantify the effect of the fuel-price shock on mobility range separately for the US and China (see ``Quantifying mobility rigidity'').
Finally, we describe the models used to disentangle the dimensions of mobility rigidity and quantify each dimension's contribution to predicting mobility rigidity (see ``Modelling the structural dimensions of mobility rigidity'').

\subsection*{Datasets}

\paragraph{Advan Research Foot Traffic.} Advan Research aggregates anonymized location data from a wide range of mobile applications. 
The data record weekly movements between census block groups (CBGs), small geographic units typically containing 600–3,000 residents, and points of interest (POIs), such as restaurants, grocery stores, and parks.
This dataset measures mobility in the US (126 weeks, January 2024 – May 2026), which provides Point of Interest (POI) visit counts and their source CBG~\cite{advan2025weeklypatterns}. 
The dataset captures approximately 1,046 billion visits across 84,444 CBGs over the study period, with an average of 891 unique POIs visited per CBG per week.
For each CBG-to-POI visit, we compute the haversine distance between the CBG centroid and the visited POI, then aggregate to the CBG–week level as the visit-count-weighted mean travel distance (km). 
After data augmentation with built environment and socioeconomic data, we remove CBGs with missing data, yielding a final analytical sample of 82,908 CBGs over 113 weeks for pre-shock baseline characterization and 9 weeks for quantifying the discontinuity effect of fuel-price shock on people's mobility range.

\paragraph{Location-based Services (LBS) Data.}
Provided by a top LBS company in China, the data record movements between subdistrict (jiedao), small administrative units typically containing 5,900--36,300 residents, and POIs.
This dataset measures mobility in China, including weekly subdistrict-level travel data (82 weeks, November 2024 – May 2026) that records POI visit counts and their source subdistrict across 337 cities in 31 provinces. 
The dataset captures approximately 651 billion visits across 39,538 subdistricts over the study period. 
For each visit, we compute the haversine distance between the subdistrict centroid and the visited POI, then aggregate to the subdistrict–week level as the visit-count-weighted mean travel distance. 
The data also provide travel mode shares for driving, walking, bus, and subway. 
After merging with built environment and socioeconomic covariates and removing subdistricts with missing data, the final analytical sample comprises 39,538 subdistricts over 70 weeks for pre-shock baseline characterization and 13 weeks for estimating the discontinuity effect of the fuel-price shock on mobility range.

\paragraph{Built environment.} For the US, the EPA Smart Location Database V3 (220,740 block groups, aggregated to CBG means) provides population density (D1B), land-use diversity (D2A\_EPHHM), and street connectivity (D3B)~\cite{epa2021sld}. 
Overture Maps provide land-use polygons from which we compute the Shannon entropy of 30 land-use classes within each CBG~\cite{overture2024}. 
Transitland provides stop-level transit data, aggregated to CBG-level stop density per 10,000 residents and classified into five groups by service frequency~\cite{transitland2024}. \par

For China, all subdistrict-level attributes are derived from the data provider's feature database, which reports aggregate counts per subdistrict at the grid-cell level (1 km by 1 km). 
Population density is defined as the number of permanent residents per grid cell, sourced from the data provider. 
Transit accessibility is captured by two variables: a subway dummy variable indicating whether at least one subway station falls within the subdistrict, and bus stop density. 
Land-use diversity is measured as the Shannon entropy of the 13 POI categories (see Supplementary Table~\ref{tab:si_poi}).
For each neighbourhood in both countries, we calculate the share of total visits to each POI category and harmonize these categories into three groups — discretionary, services, and commerce — with commerce as the reference category (Supplementary Table~\ref{tab:si_poi}).

\paragraph{Socioeconomic attributes.}
For the US, the socioeconomic context is obtained from the US Census Bureau's 2024 5-year (2020--2024) American Community Survey (ACS) estimates at the CBG level~\cite{census2024acs}. 
We use the ACS 2024 5-year estimates to extract median household income and car ownership at the CBG level in the US.
For China, subdistrict-level socioeconomic covariates are obtained from the data provider's aggregated platform data. 
Car ownership is measured by the proportion of residents who own at least one private vehicle.
Neighbourhood wealth is proxied by the share of residents classified in the highest-wealth tier, which accounts for about 5\% of the total population in China.

\paragraph{Data ethics.}
Both mobility datasets were obtained in aggregated and anonymized form. 
The US data are provided by Advan Research, which aggregates location information from mobile applications whose users have opted in to the collection of anonymous location data. 
The China data are provided by a top LBS data provider and consist of aggregated mobility statistics derived from users of its location-based services who have consented to the collection of anonymized location information. 
Neither dataset contains personally identifiable information or individual-level trajectories. 
All analyses were conducted exclusively on aggregated spatial (CBG in the US and subdistrict in China) and temporal units (weekly), and no attempt was made to identify individual users.

\subsection*{Mobility rigidity}
\paragraph{Definition.}
We examine the extent to which residents of a neighbourhood (CBG in the US or subdistrict in China) maintain their pre-shock mobility range as travel costs rise.
Here, we define mobility range as the average haversine distance between a neighbourhood's centroid and the visited POIs, representing the characteristic spatial footprint of daily mobility~\cite{liao2025uncovering}.

\paragraph{Pre-shock dimensions of adaptive capacity.}
We consider five dimensions of the pre-shock structure that predict neighbourhood-level mobility rigidity.
\textit{Price-transmission ratio (PT)} scales the global fuel-price shock to the national level, reflecting how much of the crude price increase is passed through to domestic travellers via institutional filtering. 
To quantify this, we first convert all price series to a common unit (CNY per liter) using a fixed exchange rate of 7.25 CNY/USD; for WTI crude, $P_t = \text{WTI close} / 159 \times 7.25$, where 159 is the barrel-to-litre conversion factor. 
We then normalize each series to a pre-shock index by dividing by its baseline average over 1 October 2025 -- 26 January 2026 and multiplying by 100. 
The peak percentage change is $\Delta \% P = \max_{t \in [27\text{ Jan},\ 1\text{ Jul}\ 2026]} \left( \frac{P_t}{\bar{P}_{\text{base}}} \times 100 \right) - 100$, computed separately for international crude ($\Delta \% P_{\text{crude}}$) and domestic retail gasoline ($\Delta \% P_{\text{retail}}$). 
The price-transmission ratio is then $\text{PT} = \frac{\Delta \% P_{\text{retail}}}{\Delta \% P_{\text{crude}}}$, which operates at the national level and is used to interpret cross-country differences in mobility rigidity. \par

\textit{Exposure} measures how dependent a spatial unit is on energy-intensive travel under normal conditions, quantified as the pre-shock baseline travel distance (neighbourhood-to-POI distance in km) and vehicle ownership.
Neighbourhoods where residents tend to use cars and travel longer distances are more exposed to fuel-price shocks because a given increase in fuel prices translates into a larger increase in travel costs. \par

\textit{Sensitivity} is represented by the share of essential trips and discretionary trips.
We compute two pre-shock indicators to reflect this dimension: the share of visits to discretionary and services POIs (see Supplementary Table~\ref{tab:si_poi}). \par

\textit{Absorptive capacity} (AC) represents the financial resources available to absorb higher fuel costs, measured by log median household income (USD), from ACS 2024 5-year estimates at the CBG level. 
AC in China is proxied by wealth (share of residents in the highest wealth tier), sourced from the data provider's LBS platform at the subdistrict level. \par

\textit{Substitutive capacity} (SC) describes the built environment that enables behavioural adjustment, captured through transit stop density (transit stops per 10,000 residents, winsorized at the 99th percentile), land-use diversity quantified via Overture Maps land-use polygons (30 classes including residential, commercial, industrial, park, school, hospital, religious, and others), intersected with Census 2020 CBG boundaries. 
For each CBG, the Shannon entropy of class area shares is computed as:

\begin{equation}
H_i = -\sum_{c=1}^{C} \frac{A_{ic}}{A_i} \cdot \ln\left(\frac{A_{ic}}{A_i}\right)
\end{equation}
where $C$ is the number of classes present in CBG $i$, $A_{ic}$ is the polygon area of class $c$ within the CBG, and $A_i$ is the total land area of the CBG. 
Higher $H_i$ indicates a more balanced land-use mix. 
Proximity to typical destinations is measured as the haversine distance (km) from the neighbourhood centroid to the nearest school, commercial area, and gas station.

\subsection*{Quantifying mobility rigidity}
Mobility rigidity is quantified as the neighbourhood-level shock responses via a two-level hierarchical regression discontinuity design (RDD):
\begin{equation}
\log(\text{dist}_{it}) = \beta_{0i} + \beta_{1i}\cdot\text{post}_t + \beta_{2}\cdot t + \beta_{3}\cdot\text{post}_t\times t + \text{controls} + \varepsilon_{it}
\label{eq:rdd}
\end{equation}
where $\text{dist}_{it}$ is the visit-weighted mean travel distance for neighbourhood $i$ in week $t$, $\beta_{0i}$ is the neighbourhood-specific baseline log distance, $\beta_{1i}$ captures the immediate change in log distance following the shock, $\beta_{2}$ is the general pre-shock linear time trend, and $\beta_{3}$ allows the trend to change after the shock. \par

The running variable $t$ is the number of weeks relative to the cutoff (centred at zero on 28 February 2026 for the US and 10 March for China), and $\text{post}_t$ is an indicator for weeks after $t \geq 0$. 
Controls include week-of-year and year-month fixed effects to absorb seasonality and a sampling-proxy control (log POI coverage) for data coverage. \par

At Level 2, the neighbourhood-specific coefficients are modelled as:
\begin{equation}
\beta_{0i} = \gamma_{00} + u_{0i}, \quad \beta_{1i} = \gamma_{10} + u_{1i}
\end{equation}
where $\gamma_{00}$ is the population mean baseline log distance, $\gamma_{10}$ is the population mean shock effect, and $u_{0i}$ and $u_{1i}$ are neighbourhood-specific random deviations assumed to follow a bivariate normal distribution $(u_{0i}, u_{1i}) \sim \mathcal{N}(0, \Sigma)$~\cite{cattaneo2016interpreting, cattaneo2024practical}. \par

The Best Linear Unbiased Predictor (BLUP) of $u_{1i}$ yields the neighbourhood-level deviation from the mean shock effect~\cite{raudenbush2015multisite, rhoads2016optimal}. 
By shrinking noisier estimates toward the population mean~\cite{pagan1984econometric}, the BLUP reduces the influence of neighbourhoods with few observations while preserving the spatial heterogeneity of interest. 
The neighbourhood-specific discontinuity effect is then obtained as $\hat\tau_i=\hat{\gamma}_{10}+\hat{u}_{1i}$ as the mobility rigidity.
Because the random-effect predictions are shrunk toward the population mean, each $\hat\tau_i$ carries less variance than the true effect.
We quantify this attenuation by the reliability of the BLUP, $\lambda = \tau_1^2/(\tau_1^2 + \bar v)$, where $\tau_1^2$ is the between-neighbourhood variance of the true effects and $\bar v$ is the average conditional (posterior) variance of the $\hat\tau_i$ from the fitted model. \par

The main specification uses bandwidths in weeks: $bw=4$ (US) and $bw=6$ (China, for holiday exclusion, as detailed in Supplementary Table~\ref{tab:holiday_exclusion}), with a linear time trend and neighbourhood-specific linear pre-trends, estimated via restricted maximum likelihood~\cite{bates2015fitting}. 
Robustness checks, including placebo tests and alternative bandwidth estimates~\cite{cattaneo2016interpreting}, confirm the main results (Supplementary Tables 3–6). 
The final sample comprises 82,908 (US) and 39,538 (China) neighbourhoods with BLUP estimates. 
We decompose the variance of $\hat{\tau}_i$ into between- and within-city components by one-way analysis of variance with cities as groups, using administrative boundaries in China and Core-based Statistical Area (CBSA) in the US. \par

\subsection*{Modelling the structural dimensions of mobility rigidity}

We estimate four progressive Ordinary Least Squares (OLS) models of increasing complexity to quantify the incremental explanatory power of each pre-shock structural dimension:

\begin{align}
  \text{G1:} \quad \hat{\tau}_i &= \alpha + \boldsymbol{\beta}_1' \, \mathbf{X}^{\text{Exp}}_i + \varepsilon_i \\
  \text{G2:} \quad \hat{\tau}_i &= \alpha + \boldsymbol{\beta}_1' \, \mathbf{X}^{\text{Exp}}_i + \boldsymbol{\beta}_2' \, \mathbf{X}^{\text{SC}}_i + \varepsilon_i \\
  \text{G3:} \quad \hat{\tau}_i &= \alpha + \boldsymbol{\beta}_1' \, \mathbf{X}^{\text{Exp}}_i + \boldsymbol{\beta}_2' \, \mathbf{X}^{\text{SC}}_i + \boldsymbol{\beta}_3' \, \mathbf{X}^{\text{AC}}_i + \varepsilon_i \\
  \text{G4:} \quad \hat{\tau}_i &= \alpha + \boldsymbol{\beta}_1' \, \mathbf{X}^{\text{Exp}}_i + \boldsymbol{\beta}_2' \, \mathbf{X}^{\text{SC}}_i + \boldsymbol{\beta}_3' \, \mathbf{X}^{\text{AC}}_i + \boldsymbol{\beta}_4' \, \mathbf{X}^{\text{S}}_i + \varepsilon_i
\end{align}\label{eq:ols}

\noindent where $\mathbf{X}^{\text{Exp}} = [\text{bsl}_i, \text{veh}_i]'$ denotes the Exposure dimension (baseline travel distance and vehicle ownership), and $\mathbf{X}^{\text{SC}}$, $\mathbf{X}^{\text{AC}}$, and $\mathbf{X}^{\text{S}}$ denote the Substitutive capacity (density, transit, land-use diversity, proximity), Absorptive capacity (income), and Sensitivity (discretionary and services POI shares) variable vectors, respectively. 
The $\Delta R^2_{\mathrm{BLUP}}$ between successive models isolates the marginal contribution of each variable group after controlling for all prior groups. \par

At this stage, the dependent variable $\hat\tau_i$ is a shrunk estimate with a defined reliability $\lambda$, the attainable $R^2$ of these regressions is attenuated: the observed $R^2_{\mathrm{BLUP}}$ is $\lambda$ times the value that would be obtained with the true effects.
Therefore, we report a reliability-corrected $R^2 = R^2_{\mathrm{BLUP}}/\lambda$, which removes the variance lost to shrinkage and represents the estimated ceiling on the explanatory power of OLS models. \par

The progressive $R^2$ approach reveals the contribution of variable \textit{groups} but does not isolate individual predictors, because the marginal $\Delta R^2$ of a variable depends on which other variables are already in the model. 
To decompose the full-model $R^2$ into unique, additive, order-independent contributions of each predictor, we use the Lindeman--Merenda--Gold (LMG) method~\cite{lindeman1980introduction, gromping2007relative}.

The LMG value for predictor $x_j$ is defined as the average increase in $R^2$ when $x_j$ is added, computed over all possible orderings of the $p$ predictors:

\begin{equation}
  \text{LMG}(x_j) = \frac{1}{p!} \sum_{\pi \in \mathcal{S}_p} \left[ R^2(S_\pi^{(j)} \cup \{x_j\}) - R^2(S_\pi^{(j)}) \right]
\end{equation}

\noindent where $\mathcal{S}_p$ is the set of all $p!$ permutations of the $p$ predictors, $S_\pi^{(j)}$ is the set of predictors that precede $x_j$ in permutation $\pi$, and $R^2(S)$ is the coefficient of determination from regressing $\hat{\tau}_i$ on the predictors in set $S$. The LMG values sum exactly to the full-model $R^2$:

\begin{equation}
  \sum_{j=1}^{p} \text{LMG}(x_j) = R^2_{\text{full}}
\end{equation}

We report LMG values as percentages of the full-model $R^2$ (i.e., $\text{LMG}_\%(x_j) = \text{LMG}(x_j) / R^2_{\text{full}} \times 100$), which sums to 100\% and represents each predictor's share of explained variance. \par

To summarise the LMG results at the conceptual level, we aggregate individual LMG values within each dimension of the mobility rigidity framework:

\begin{equation}
  \text{LMG}_{\text{dim}}(D) = \sum_{j: x_j \in D} \text{LMG}(x_j)
\end{equation}

\noindent where $D \in \{\text{Exposure, Sensitivity, Absorptive capacity, Substitutive capacity}\}$. 
This aggregation yields the share of explained variance attributable to each structural dimension. 
The price-transmission ratio (PT) operates at the national level and is therefore excluded from the neighbourhood-level LMG decomposition.

\section*{Declarations}
\textbf{Data Availability}.
Advan Research Foot Traffic Weekly Patterns data are available via the Dewey platform under an academic license.
EPA Smart Location Database V3 is publicly available at \url{https://www.epa.gov/smart-growth/smart-location-mapping}.
ACS 2024 5-year estimates are available via \url{https://www.census.gov/programs-surveys/acs}.
Transitland stop data is available at \url{https://www.transit.land/}.
Mobility data for China were obtained from a commercial location-based services provider under a restricted research agreement. 
Requests for access should be directed to the corresponding author (X.M.).

\noindent\textbf{Code Availability}.
The code to reproduce the analysis is available on \href{https://github.com/ZHANGZIHAO2390/oil-shock-mobility-adaptation}{GitHub}.

\noindent\textbf{Acknowledgements}.
Access to the Dewey data platform was supported by the Swedish Research Council (Project Number 2022-06215).
X.M. acknowledges support from the National Natural Science Foundation of China (W2611102).

\noindent\textbf{Author Contributions}.
Y.L. and Z.Z. designed the study. 
Z.Z. performed the preliminary exploratory analysis. 
Z.Z. and Y.Z. performed the analysis under the supervision of X.M. and Y.L.
All authors contributed to the manuscript writing.

\noindent\textbf{Competing Interests}.
The authors declare no competing interests.

\renewcommand\refname{References}
\bibliography{bibliography}

\newpage
\begin{appendices}

\renewcommand{\thefigure}{\arabic{figure}}
\renewcommand{\thetable}{\arabic{table}}
\renewcommand{\figurename}{Supplementary Figure}
\renewcommand{\tablename}{Supplementary Table}
\setcounter{figure}{0}
\setcounter{table}{0}

\begin{center}
    {\LARGE \bfseries Supplementary Information\\}
    \vspace{1cm}
    {\LARGE \bfseries Unequal urban capacities for mobility adaptation under fuel-price shocks}
\end{center}

\section*{Fuel-price transmission ratio in the US and China}

\begin{figure}[h]
    \centering
    \includegraphics[width=0.6\linewidth]{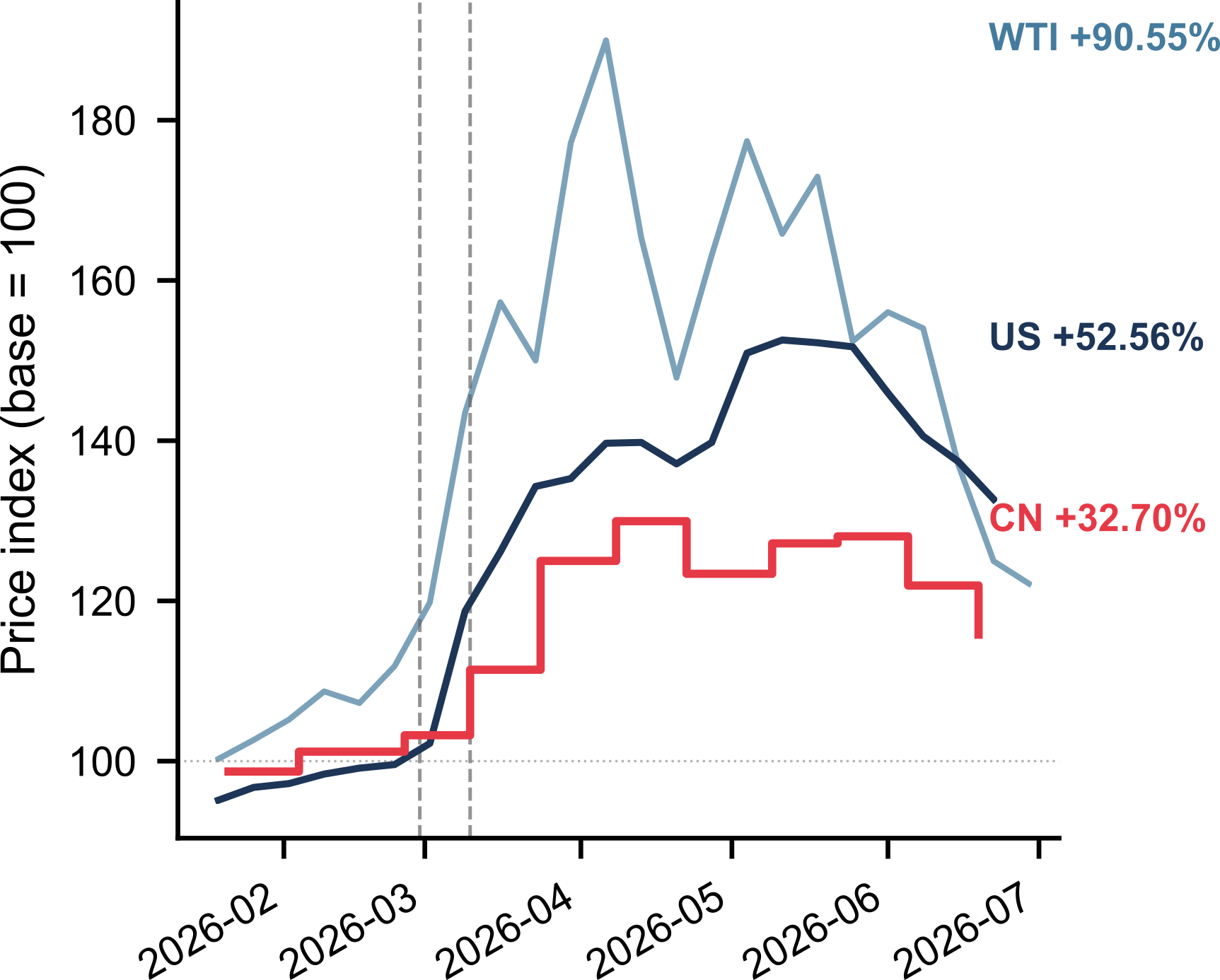}
    \caption{Crude oil (WTI) and retail gasoline price series for the US and China, January -- June 2026. Vertical dashed lines indicate the respective price breakpoints.}
    \label{fig:si_oil_price}
\end{figure}

\begin{table}[h]
\centering
\caption{Price-transmission ratio calculation details. 
Baseline period: 1 October 2025 -- 26 January 2026; peak window: 27 January -- 1 July 2026.}
\label{tab:si_pt}
\begin{tabular}{lccc}
\toprule
\textbf{Component} & \textbf{WTI crude} & \textbf{US retail} & \textbf{China retail} \\
\midrule
Baseline average & \$59.07/bbl & \$2.95/gal & 7,782 yuan/tonne \\
Peak value & \$112.56/bbl (6 Apr) & \$4.50/gal (11 May) & 10,325 yuan/tonne (8 Apr) \\
Peak change ($\Delta\%P$) & 90.55\% & 52.56\% & 32.7\% \\
PT ($\Delta\%P_{\text{retail}}/\Delta\%P_{\text{crude}}$) & — & 0.58 & 0.36 \\
\bottomrule
\end{tabular}
\end{table}

Price series are later converted to a common unit (CNY per liter) using a fixed exchange rate of 7.25 CNY/USD. 
WTI crude is converted as $P_t = \text{WTI close} / 159 \times 7.25$ (barrel-to-litre factor: 159). US retail gasoline is converted as $P_t = \text{GASREGW} / 3.785 \times 7.25$ (gallon-to-litre factor: 3.785). 
Chinese retail gasoline prices are obtained directly from the NDRC 92-octane price in CNY per litre.

\subsection*{Holiday week exclusion}

Because Chinese public holidays generate large, predictable departures from normal travel patterns, we exclude weeks containing major national holidays from the estimation sample. Holiday timing follows the lunar calendar and therefore shifts across ISO weeks between years; we apply year-specific exclusion rules to ensure consistent treatment across the 2025 and 2026 portions of the panel. Supplementary Table~\ref{tab:holiday_exclusion} documents the full set of excluded weeks.

\paragraph{Spring Festival (Chinese New Year).}
Spring Festival is the largest annual disruption to urban mobility in China, producing a 2--4 week period of elevated long-distance travel (chunyun) followed by suppressed local activity. In 2025, the Lunar New Year falls on 29 January (ISO week 5); we exclude ISO weeks 4--7 (20 Jan -- 16 Feb), covering the pre-holiday travel rush through the post-holiday return. In 2026, the Lunar New Year falls on 17 February (ISO week 8); we exclude ISO weeks 7--9 (9 Feb -- 1 Mar), a slightly shorter window reflecting its later calendar position and avoiding overlap with the estimation window.

\paragraph{Other statutory holidays.}
We additionally exclude weeks containing four other national holidays that produce measurable travel anomalies:
\begin{itemize}
  \item \textit{New Year's Day} (1 January): ISO week 1 in both years.
  \item \textit{Qingming Festival}: ISO week 14 in 2025; ISO weeks 14--15 in 2026 (the holiday falls on a weekend in 2026, extending the compensatory leave into the following week).
  \item \textit{Labor Day} (1 May): ISO weeks 18--19 in both years (the 5-day holiday plus adjacent weekend spans two weekly observation periods).
  \item \textit{Dragon Boat Festival} (Duanwu): ISO weeks 22--23 in 2025; ISO week 25 in 2026 (the lunar date shifts from late May to mid-June).
\end{itemize}

After exclusion, the estimation sample retains 45 valid weeks in 2025 (of 53 total) and 44 valid weeks in 2026.
Of the 23 available weeks through the data end-date, we have 13 consecutive valid weeks within the primary estimation window ($bw = 6$, centred on 10 March 2026).

\begin{table}[h]
\centering
\caption{Holiday week exclusion schedule for China. ISO weeks are excluded on a per-year basis to account for lunar calendar shifts.}
\label{tab:holiday_exclusion}
\begin{tabular}{llll}
\hline
Holiday & 2025 (ISO weeks) & 2026 (ISO weeks) & Rationale \\
\hline
\textit{Spring Festival} & 4, 5, 6, 7 & 7, 8, 9 & Chunyun travel rush \\
 & (20 Jan -- 16 Feb) & (9 Feb -- 1 Mar) & and post-holiday return \\[4pt]
\textit{New Year's Day} & 1 & 1 & Statutory holiday \\
 & (30 Dec -- 5 Jan) & (29 Dec -- 4 Jan) & \\[4pt]
\textit{Qingming} & 14 & 14, 15 &  3-day holiday \\
 & (31 Mar -- 6 Apr) & (30 Mar -- 12 Apr) &  \\[4pt]
\textit{Labor Day} & 18, 19 & 18, 19 & 5-day holiday spans \\
 & (28 Apr -- 11 May) & (27 Apr -- 10 May) & two observation weeks \\[4pt]
\textit{Dragon Boat} & 22, 23 & 25 & Lunar date shifts \\
 & (26 May -- 8 Jun) & (15 Jun -- 21 Jun) & from late May to mid-Jun \\[4pt]
\hline
\multicolumn{4}{l}{\textit{Total excluded:} 10 weeks (2025), 9 weeks (2026)} \\
\hline
\end{tabular}
\begin{flushleft}
\footnotesize\textit{Notes:} Date ranges indicate the Monday--Sunday span of each ISO week. The 2026 data end at ISO week 23 (week starting 1 June 2026). The RDD running variable $t$ is built on the post-exclusion valid-week sequence; excluded weeks do not occupy positions. $bw = 6$ yields 6 pre-cutoff + 7 post-cutoff valid weeks = 13 (ISO weeks 3--6, 10--13, 16--17, 20--22).
\end{flushleft}
\end{table}

\section*{Bandwidth sensitivity and placebo tests}
\paragraph{United States.}The hierarchical RDD is estimated across bandwidths $bw=2$ to $bw=8$ (Supplementary Table~\ref{tab:si_us_bandwidth}). 
The estimated shock response is statistically significant at $bw=2-4$ and $bw=8$, with magnitudes decreasing from -23.6\% to -10.3\% as bandwidth widens. 
The intermediate bandwidth ($bw=6$) shows no significant effect, because this wider pre-window extends back to the weeks immediately following the late-January geopolitical signal, when prices had already begun rising, and travel was already adjusting. 
Part of the contraction is therefore absorbed into the pre-shock baseline rather than attributed to the post-shock discontinuity.
A placebo test that assigns the pseudo-breakpoint to 28 February 2025 (one year before the actual shock) produces no statistically significant effect across any bandwidth (all $p > 0.10$).

\begin{table}[h]
\centering
\caption{Bandwidth sensitivity and placebo test results: United States. 
Dependent variable: $\log(\text{dist}_{it})$ at the neighbourhood--week level. 
Standard errors clustered at the state level.
$^{***}p<0.01$, $^{**}p<0.05$, $^{*}p<0.1$.}
\label{tab:si_us_bandwidth}
\begin{tabular}{lcccc}
\toprule
\multirow{2}{*}[-2pt]{\textbf{bw}} & \multicolumn{2}{c}{\textbf{Main specification}} & \multicolumn{2}{c}{\textbf{Placebo (2025-02-28)}} \\
\cmidrule{2-5}
 & $\hat\tau$ & Effect (\%) & $\hat\tau$ & $p$-value \\
\midrule
2  & $-$0.270$^{***}$ & $-$23.62\% & $+$0.006 & 0.804 \\
3  & $-$0.247$^{***}$ & $-$21.85\% & $+$0.003 & 0.925 \\
4  & $-$0.211$^{***}$ & $-$19.00\% & $+$0.002 & 0.946 \\
6  & $+$0.025 & $+$2.49\%  & $+$0.003 & 0.912 \\
8  & $-$0.108$^{**}$  & $-$10.27\% & $+$0.001 & 0.979 \\
\bottomrule
\end{tabular}
\end{table}

\paragraph{China.} The same analysis is repeated for the Chinese sample with bandwidths $bw=4$ to $bw=8$, using a pseudo-breakpoint of 10 March 2025 (Supplementary Table~\ref{tab:si_cn_bandwidth}). The estimated effects are stable across bandwidths $bw=6$ to $bw=8$, and the placebo test confirms that the results are not driven by seasonal patterns in spring travel.

\begin{table}[h]
\centering
\caption{Bandwidth sensitivity and placebo test results: China. Dependent variable: $\log(\text{dist}_{it})$ at the subdistrict--week level.
$^{***}p<0.01$, $^{**}p<0.05$, $^{*}p<0.1$.}
\label{tab:si_cn_bandwidth}
\begin{tabular}{lcccc}
\toprule
\multirow{2}{*}[-2pt]{\textbf{bw}} & \multicolumn{2}{c}{\textbf{Main specification}} & \multicolumn{2}{c}{\textbf{Placebo (2025-03-10)}} \\
\cmidrule{2-5}
 & $\hat\tau$ & Effect (\%) & $\hat\tau$ & $p$-value \\
\midrule
4  & $-$0.001$^{*}$ & $-$0.13\% & $+$0.001 & 0.847 \\
5  & $-$0.007$^{***}$ & $-$0.71\% & $-$0.003 & 0.432 \\
6  & $-$0.024$^{***}$ & $-$2.33\% & $+$0.001 & 0.881 \\
7  & $-$0.024$^{***}$ & $-$2.35\% & $-$0.003 & 0.372 \\
8  & $-$0.024$^{***}$ & $-$2.34\% & $-$0.002 & 0.586 \\
\bottomrule
\end{tabular}
\end{table}

\section*{POI classification mapping}

Supplementary Table~\ref{tab:si_poi} maps the POI categories used in each country to the three functional groups employed in the progressive OLS models. 
In both countries, non-commercial NAICS categories (manufacturing, construction, agriculture, etc.) are excluded before computing visit shares.

\begin{table}[h]
\centering
\caption{POI classification mapping for the US and China. 
Commerce is the omitted reference category in the progressive OLS models.}

\label{tab:si_poi}
\begin{tabularx}{\textwidth}{@{}lXX@{}}
\toprule
\textbf{Group} & \textbf{US (NAICS)} & \textbf{China (categories)} \\
\midrule
\textbf{Discretionary} & Arts, Entertainment, and Recreation (71) & Dining, sports, accommodation \\
 & Accommodation and Food Services (72) & Scenic spots \\
\midrule
\textbf{Services} & Other Services (81) & Education, medical, Public Administration \\
 & Public Administration (92) & Finance, corporate services \\
 & Educational Services (61), Professional, Scientific, and Technical Services (54--56) & Personal services, auto services \\
 & Finance and Insurance (52) & \\
\midrule
\textbf{Commerce (ref.)} & Retail Trade (44--45) & Shopping \\
 & Transportation and Warehousing (48--49) & Business \& residential, transport \\
\bottomrule
\end{tabularx}
\end{table}

\section*{Robustness of neighbourhood-level treatment effects to shrinkage}

The obtained neighbourhood-level effects are from a hierarchical model as best linear unbiased predictions (BLUPs), and we verify that this gradient is not an artifact of shrinkage. 
Specifically, we test whether noisier units are pulled disproportionately toward the global mean in a way that correlates with absorptive capacity.
We test this in three ways: (i) whether the posterior standard error of $\hat{\tau}_i$ correlates with income; (ii) whether an un-shrunk estimator (the simple pre--post difference of residuals, hereafter ``simple difference'') reproduces the gradient; and (iii) whether the gradient survives within strata of potential confounders.

\paragraph{United States.} In the US, each neighbourhood contributes a comparable number of observations to the estimating sample, and the posterior standard error of $\hat{\tau}_i$ is nearly constant across income quartiles (median 0.318 in all four quartiles). 
Consequently, the correlation between $\hat{\tau}_i$ standard errors and log income is negligible ($r = 0.013$), and the correlation between actual shrinkage and log income is weak ($r = 0.108$), where actual shrinkage is defined as the absolute difference between the un-shrunk simple-difference estimate and the BLUP ($|\hat{\tau}_i^{\mathrm{simple}} - \hat{\tau}_i^{\mathrm{BLUP}}|$). 
These correlations fall below the thresholds at which shrinkage would be considered systematically income-dependent.

Supplementary Table~\ref{tab:shrinkage_us} compares the absorptive-capacity gradient, defined as the difference in mean treatment effect between the highest ($Q_4$) and lowest ($Q_1$) income quartiles, under the BLUP and under un-shrunk estimators. 
The simple-difference estimator, which does not borrow information across neighbourhoods, reproduces the gradient with the same sign and a larger magnitude ($+0.096$ versus $+0.053$). 
Precision-weighting the simple-difference estimates by the inverse of their variances yields a gradient of $+0.046$, still of the same sign. 

The income gradient also persists within all four strata of baseline visit frequency ($+0.031$ to $+0.086$), confirming that it is not an artefact of differential shrinkage across frequency groups. 
Shrinkage, therefore, compresses rather than generates the reported income gradient in the US.

\begin{table}[htbp]
\centering
\caption{United States: absorptive-capacity gradient under alternative estimators.}
\label{tab:shrinkage_us}
\begin{tabular}{lccc}
\toprule
Estimator & $\hat{\tau}_{Q_1}$ & $\hat{\tau}_{Q_4}$ & $Q_4 - Q_1$ \\
\midrule
BLUP & $-0.233$ & $-0.180$ & $+0.053$ \\
Simple difference (raw) & $-0.025$ & $+0.071$ & $+0.096$ \\
Simple difference (precision-weighted) & $-0.135$ & $-0.089$ & $+0.046$ \\
\bottomrule
\end{tabular}
\end{table}

\paragraph{China.} The Chinese panel has a balanced structure: every subdistrict contributes exactly thirteen weekly observations (six pre-treatment and seven post-treatment), so differential shrinkage cannot arise from differences in sample size across units. 
Consistent with this, sample size is uncorrelated with car ownership ($r = 0.017$) and with the share of high-wealth households ($r = 0.003$).

The un-shrunk simple-difference estimator reproduces the absorptive-capacity gradient with the same sign and a larger magnitude than the BLUP ($+0.057$ versus $+0.041$; Supplementary Table~\ref{tab:shrinkage_cn}). 
The gradient also persists within all four strata of baseline visit frequency ($+0.017$ to $+0.116$), and the ratio of BLUP to simple-difference variances is stable across frequency groups (0.25--0.32), providing no evidence of frequency-dependent shrinkage. 
The correlation between the simple-difference and BLUP estimates is high ($r = 0.880$), indicating that shrinkage compresses noise without reordering units.

\begin{table}[htbp]
\centering
\caption{China: absorptive-capacity gradient under alternative estimators.}
\label{tab:shrinkage_cn}
\begin{tabular}{lccc}
\toprule
Estimator & $\hat{\tau}_{Q_1}$ & $\hat{\tau}_{Q_4}$ & $Q_4 - Q_1$ \\
\midrule
BLUP & $-0.045$ & $-0.004$ & $+0.041$ \\
Simple difference (raw) & $-0.030$ & $+0.027$ & $+0.057$ \\
\bottomrule
\end{tabular}
\end{table}

\paragraph{Cross-country comparison.} The two countries differ in the source of potential differential shrinkage. 
In the US, shrinkage could, in principle, vary with neighbourhood-level sample size; in China, it cannot because the panel is balanced. 
In both settings, however, the unshrunken estimator reproduces the gradient with the same sign and larger or comparable magnitude. 
The above analysis suggests that the absorptive-capacity gradient reported in the main text is not a statistical artifact of BLUP shrinkage.

\section*{External validation of estimated shock effects}
The shock produced plausible but sharply different national responses: the implied mobility-price elasticity was approximately -0.50 in the US (-19.0\% mobility change following a 52.6\% price increase) and -0.08 in China (-2.33\% following a 32.7\% increase).
The US estimate falls within the -0.29 to -0.61 range reported in high-frequency studies, whereas the Chinese estimate is below the -0.20 to -0.50 range reported for gasoline demand~\cite{levin2017high,goodwin2004elasticities,graham2002demand,lin2013elasticity}.
China's smaller response is consistent with lower car ownership, the inclusion of travellers using all transport modes, and weaker price transmission under administrative pricing.

\section*{Reduced driving distance and modal shift in China}

The hierarchical RDD is estimated separately for six outcome variables at the main specification bandwidth ($bw = 6$). 
Supplementary Table~\ref{tab:multi_outcome} reports the population-mean treatment effect ($\hat{\tau}$), the implied percentage change ($e^{\hat{\tau}} - 1$), and the share of neighbourhoods with negative BLUP treatment effects (i.e., the proportion responding in the expected direction).

\begin{table}[htbp]
\centering
\caption{Multi-outcome hierarchical RDD results for China ($bw = 6$, cutoff = 2026-03-10).
Each row is a separate hierarchical RDD estimation. 
Non-motorized = walking + cycling (bicycle and e-bike). 
$\Delta\%$ denotes $e^{\hat{\tau}} - 1$. 
\% Neg.\ BLUP indicates the proportion of neighbourhoods with $\hat{\tau}_i < 0$. 
$^a$ All estimates with $p < 0.001$.}
\label{tab:multi_outcome}
\begin{tabular}{l c c c c}
\hline
Outcome & $\hat{\tau}^a$ & $\Delta\%$ & $p$-value & \% Neg.\ BLUP \\
\hline
\multicolumn{5}{l}{\textit{Aggregate}} \\
\ \ Travel distance  & $-$0.0236 & $-$2.33\% & $<$0.001 & 72.1\% \\
\ \ Visit frequency  & $-$0.0181 & $-$1.79\% & $<$0.001 & 95.0\% \\
\hline
\multicolumn{5}{l}{\textit{Mode share}} \\
\ \ Driving          & $-$0.0089 & $-$0.89\% & $<$0.001 & 80.8\% \\
\ \ Non-motorized    & $+$0.0122 & $+$1.23\% & $<$0.001 & \ \ 9.8\% \\
\ \ Subway           & $-$0.0083 & $-$0.83\% & $<$0.001 & 97.8\% \\
\ \ Bus              & $-$0.0268 & $-$2.65\% & $<$0.001 & 66.3\% \\
\hline
\end{tabular}
\end{table}

The fuel-price shock reduced both total trip volume ($-$1.79\%) and driving share ($-$0.89\%), while non-motorized share increased ($+$1.23\%). Subway ($-$0.83\%) and bus ($-$2.65\%) shares also declined. Cross-neighbourhood BLUP correlations indicate that neighbourhoods where driving share fell most are the same neighbourhoods where non-motorized share rose most ($r = -0.65$), consistent with mode substitution. In contrast, the driving BLUP and the bus BLUP are \textit{positively} correlated ($r = +0.25$), indicating that driving and bus co-vary under a common factor (trip suppression) rather than substitute for each other.

This substitution pattern is consistent with the economics of marginal trip adjustment in response to a moderate fuel-price shock. 
The overall effect size is small ($-$0.89\% driving share), implying that only the most price-elastic trips are affected --- perhaps predominantly short, marginal trips for which walking and cycling provide zero-cost, door-to-door alternatives. 
For such trips, public transit offers no cost advantage: the marginal fuel cost of a 1--3~km drive is approximately \textyen 0.6--1.9 (at 8~L/100~km and \textyen 8/L), comparable to or below typical bus fares (\textyen 2 per trip), while walking or cycling the same distance takes 10--20 minutes with no monetary cost. The suppressed marginal driving trips, therefore, naturally flow to non-motorized modes rather than to public transit.

A quartile analysis of BLUP heterogeneity further supports this interpretation (Supplementary Table~\ref{tab:quartile_gradient}). Sorting neighbourhoods by the magnitude of their non-motorized share increase reveals a monotonic gradient: neighbourhoods with the largest non-motorized gains (Q4: $+$2.84\%) experienced the steepest distance decline ($-$5.76\%) and driving reduction ($-$2.47\%), while those with the smallest gains (Q1: $-$0.22\%) saw distance slightly increase ($+$0.28\%).

\begin{table}[htbp]
\centering
\caption{Quartile gradient: non-motorized substitution intensity and distance reduction.
neighbourhoods sorted into quartiles by non-motorized BLUP treatment effect. 
Values are mean $\Delta\% = (e^{\bar{\tau}} - 1) \times 100$ within each quartile.}
\label{tab:quartile_gradient}
\begin{tabular}{l c c c}
\hline
Quartile & Non-motorized $\Delta\%$ & Distance $\Delta\%$ & Driving $\Delta\%$ \\
\hline
Q1 (least) & $-$0.22 & $+$0.28 & $+$0.52 \\
Q2         & $+$0.88 & $-$1.00 & $-$0.52 \\
Q3         & $+$1.43 & $-$2.73 & $-$1.07 \\
Q4 (most)  & $+$2.84 & $-$5.76 & $-$2.47 \\
\hline
\end{tabular}
\end{table}

Of the total distance decline of $-$2.33\%, approximately $-$1.64\% is attributable to compositional mode shift (with the driving-share decline contributing the largest component at $-$0.97\%), and the remaining $-$0.69\% reflects trip suppression --- the cancellation of longer trips.

\section*{Multicollinearity test of the variables in the full model (G4)}

To assess whether the coefficient signs and significance levels in the full model (G4) are affected by multicollinearity, we compute variance inflation factors (VIF) for all predictors (Supplementary Table~\ref{tab:si_vif}). 
All VIF values are below 5, indicating no severe multicollinearity.

\begin{table}[h]
\centering
\caption{Variance inflation factors for the full model G4.}
\label{tab:si_vif}
\small
\begin{tabular}{llcc}
\toprule
\textbf{Dimension} & \textbf{Predictor} & \textbf{VIF (US)} & \textbf{VIF (China)} \\
\midrule
\multirow{2}{*}{Exposure}
    & Baseline distance       & 1.75 & 1.23 \\
    & Vehicle/Car ownership   & 2.20 & 1.59 \\
\addlinespace
\multirow{1}{*}{Absorptive capacity}
    & Income/Wealth           & 1.56 & 2.43 \\
\addlinespace
\multirow{6}{*}{Substitutive capacity}
    & Transit density         & 1.14 & 3.90 \\
    & Land-use diversity      & 1.05 & 1.45 \\
    & Population density      & 1.85 & 4.51 \\
    & School distance         & 2.39 & N/A  \\
    & Commercial distance     & 2.23 & N/A  \\
    & Subway                  & N/A  & 2.16 \\
\addlinespace
\multirow{2}{*}{Sensitivity}
    & Discretionary POI share & 1.72 & 1.13 \\
    & Services POI share      & 1.31 & 1.35 \\
\bottomrule
\end{tabular}
\end{table}

\section*{Progressive regression models outcomes}

Supplementary Tables~\ref{tab:si_g1g4}--\ref{tab:si_g1g4_cn} reports the full regression results for the four progressive OLS models (G1--G4).

\begin{table}[h]
\centering
\caption{Progressive OLS regression results for the US. Dependent variable: $\hat\tau_i$. Standard errors are HC3.}
\label{tab:si_g1g4}
\small
\begin{tabular}{lcccc}
\toprule
\textbf{Predictor} & \textbf{G1} & \textbf{G2} & \textbf{G3} & \textbf{G4} \\
\midrule
Baseline distance      & $-$0.0276 (0.0020)$^{***}$ & $-$0.0243 (0.0025)$^{***}$ & $-$0.0302 (0.0026)$^{***}$ & $-$0.0330 (0.0027)$^{***}$ \\
Vehicle ownership      & $+$0.0352 (0.0020)$^{***}$ & $+$0.0334 (0.0025)$^{***}$ & $+$0.0240 (0.0028)$^{***}$ & $+$0.0276 (0.0029)$^{***}$ \\
Pop.\ density          & ---                         & $-$0.0037 (0.0022)$^{*}$  & $-$0.0088 (0.0027)$^{***}$ & $-$0.0118 (0.0030)$^{***}$ \\
Land-use diversity     & ---                         & $-$0.0048 (0.0016)$^{***}$ & $-$0.0050 (0.0016)$^{***}$ & $-$0.0060 (0.0016)$^{***}$ \\
School distance        & ---                         & $-$0.0038 (0.0022)$^{*}$  & $+$0.0017 (0.0023)        & $+$0.0010 (0.0023) \\
Commercial distance    & ---                         & $-$0.0054 (0.0021)$^{**}$ & $-$0.0041 (0.0021)$^{*}$  & $-$0.0033 (0.0021) \\
Transit density        & ---                         & $-$0.0039 (0.0018)$^{**}$ & $-$0.0035 (0.0018)$^{**}$ & $-$0.0040 (0.0018)$^{**}$ \\
Income                 & ---                         & ---                        & $+$0.0187 (0.0020)$^{***}$ & $+$0.0156 (0.0021)$^{***}$ \\
Disc.\ POI share       & ---                         & ---                        & ---                        & $+$0.0093 (0.0022)$^{***}$ \\
Services POI share     & ---                         & ---                        & ---                        & $-$0.0046 (0.0020)$^{**}$ \\
\midrule
$R^2$                  & 0.0094                      & 0.0100                     & 0.0120                     & 0.0127 \\
$\Delta R^2$           & ---                         & $+$0.0006                  & $+$0.0020                  & $+$0.0008 \\
$N$                    & 82,908                      & 82,908                     & 82,908                     & 82,908 \\
\bottomrule
\end{tabular}
\end{table}

\begin{table}[h]
\centering
\caption{Progressive OLS regression results for China. Dependent variable: $\hat\tau_i$. Standard errors are HC3.}
\label{tab:si_g1g4_cn}
\small
\begin{tabular}{lcccc}
\toprule
\textbf{Predictor} & \textbf{G1} & \textbf{G2} & \textbf{G3} & \textbf{G4} \\
\midrule
Baseline distance      & $-$0.0053 (0.0004)$^{***}$ & $-$0.0054 (0.0004)$^{***}$ & $-$0.0057 (0.0004)$^{***}$ & $-$0.0072 (0.0004)$^{***}$ \\
Car ownership          & $+$0.0138 (0.0002)$^{***}$ & $+$0.0137 (0.0003)$^{***}$ & $+$0.0134 (0.0003)$^{***}$ & $+$0.0131 (0.0003)$^{***}$ \\
Pop.\ density          & ---                         & $+$0.0020 (0.0002)$^{***}$ & $+$0.0008 (0.0002)$^{***}$ & $+$0.0003 (0.0002) \\
Subway                 & ---                         & $+$0.0056 (0.0006)$^{***}$ & $-$0.0007 (0.0007)        & $+$0.0007 (0.0007) \\
Transit density        & ---                         & $+$0.0007 (0.0003)$^{***}$ & $+$0.0008 (0.0003)$^{***}$ & $-$0.0007 (0.0002)$^{***}$ \\
Land-use diversity     & ---                         & $-$0.0031 (0.0004)$^{***}$ & $-$0.0033 (0.0004)$^{***}$ & $-$0.0032 (0.0004)$^{***}$ \\
Wealth                 & ---                         & ---                        & $+$0.0033 (0.0002)$^{***}$ & $+$0.0026 (0.0002)$^{***}$ \\
Disc.\ POI share       & ---                         & ---                        & ---                        & $+$0.0079 (0.0004)$^{***}$ \\
Services POI share     & ---                         & ---                        & ---                        & $-$0.0034 (0.0003)$^{***}$ \\
\midrule
$R^2$                  & 0.1483                      & 0.1575                     & 0.1602                     & 0.2016 \\
$\Delta R^2$           & ---                         & $+$0.0092                  & $+$0.0027                  & $+$0.0414 \\
$N$                    & 39,538                      & 39,538                     & 39,538                     & 39,538 \\
\bottomrule
\end{tabular}
\end{table}

To complement the coefficient-level analysis, Supplementary Table~\ref{tab:si_lmg} reports the LMG decomposition, which partitions the full-model $R^2$ into order-independent contributions of each predictor, with bootstrap 95\% confidence intervals (200 iterations). 

\begin{table}[h]
\centering
\caption{LMG variable importance (\% of $R^2$), BLUP basis.}
\label{tab:si_lmg}
\small
\begin{tabular}{lrr}
\toprule
\textbf{Dimension / Predictor} & \textbf{US [95\% CI]} & \textbf{China [95\% CI]} \\
\midrule
\multicolumn{3}{l}{\textit{Exposure}} \\
\quad Baseline distance   & 27.28 [19.66, 33.67] & 11.6 [9.6, 14.0] \\
\quad Vehicle/Car own.    & 32.95 [25.92, 40.50] & 42.6 [39.9, 45.3] \\
\quad Total               & 60.23                 & 54.2 \\
\multicolumn{3}{l}{\textit{Substitutive}} \\
\quad Transit density     & 1.61 [0.39, 3.90]     & 4.9 [4.5, 5.3] \\
\quad Land-use diversity  & 1.61 [0.36, 3.71]     & 2.5 [2.4, 2.7] \\
\quad Pop. density        & 9.04 [5.88, 12.87]    & 3.7 [3.5, 4.0] \\
\quad School distance     & 2.33 [1.53, 3.89]     & --- \\
\quad Commercial distance & 3.78 [2.18, 5.81]     & --- \\
\quad Subway              & ---                   & 3.3 [3.1, 3.7] \\
\quad Total               & 18.37                 & 14.4 \\
\multicolumn{3}{l}{\textit{Absorptive}} \\
\quad Income/Wealth       & 18.48 [13.05, 25.24]  & 5.9 [5.5, 6.3] \\
\quad Total               & 18.48                 & 5.9 \\
\multicolumn{3}{l}{\textit{Sensitivity}} \\
\quad Disc. POI share     & 2.23 [1.51, 3.82]     & 23.6 [20.9, 26.2] \\
\quad Services POI share  & 0.69 [0.40, 1.96]     & 1.8 [1.7, 2.2] \\
\quad Total               & 2.92                  & 25.4 \\
\bottomrule
\end{tabular}
\end{table}

\end{appendices}

\end{document}